\documentclass[aps,prl,twocolumn,superscriptaddress,nofootinbib,floatfix]{revtex4-2}

\usepackage{graphicx}
\usepackage{amsmath}
\usepackage{amssymb}
\usepackage{xcolor}

\newcommand{\keV}{\ensuremath{\,\mathrm{keV}}}

\newcommand{\GeV}{\ensuremath{\,\mathrm{GeV}}}
\newcommand{\TeV}{\ensuremath{\,\mathrm{TeV}}}
\newcommand{\tyr}{\ensuremath{\,\mathrm{t\,yr}}}
\newcommand{\kms}{\ensuremath{\,\mathrm{km\,s^{-1}}}}
\newcommand{\ER}{\ensuremath{E_R}}
\newcommand{\mchi}{\ensuremath{m_\chi}}
\newcommand{\vmin}{\ensuremath{v_{\rm min}}}
\newcommand{\vesc}{\ensuremath{v_{\rm esc}}}

\newcommand{\mDM}{m_{\rm{DM}}}
\newcommand{\rhoDM}{\rho_{\rm{DM}}}

\newcommand{\shortsection}[1]{\emph{#1 —}}

\begin{document}

\title{Seasonal dark matter from the LUX-ZEPLIN high-energy event}

\author{Christopher McCabe}
\email{christopher.mccabe@kcl.ac.uk}
\affiliation{Department of Physics, King's College London, London WC2R 2LS, UK}

\date{\today}

\begin{abstract}
The LUX-ZEPLIN experiment has reported a single nuclear recoil at $248\pm23({\rm stat})\pm23({\rm sys})\keV$ in an extended search window. 
Inelastic dark matter, in which the scattering is endothermic, is a natural mechanism for producing a signal at such high energy. 
We fit the event to the dark matter mass and the mass splitting between two dark matter states, 
and find that the data favour a splitting pushed close to the kinematic limit set by the escape speed of the halo. 
The signal is then supplied by dark matter in the tail of the halo speed distribution.
This drives a strongly seasonal signal: annual modulation rises above $50\%$ across the favoured region of parameter space and even reaches $100\%$ for some parameters, implying the signal vanishes for part of the year. Such a strong modulation distinguishes this interpretation sharply from a non-modulating signal, and we quantify the number of events needed to establish it.
Experiments with larger target masses than LZ's, notably XLZD and PandaX, will be able to test this directly.
\end{abstract}

\maketitle

\shortsection{Introduction} The flagship results presented by direct searches for dark matter (as measured by citation count) typically concentrate on nuclear recoil energies below about $50\keV$ (e.g.,~\cite{LZ:2024zvo}),
since the expected spectrum from elastic scattering falls steeply with energy and any signal should therefore appear near the energy threshold of the detector~\cite{MarrodanUndagoitia:2015veg}. 
Over the past two decades, however, several classes of models have become established that suppress the signal at low energy and push it upward: effective field theory operators carrying explicit powers of the momentum transfer (e.g.,~\cite{Chang:2009yt}), and inelastic scattering~\cite{Tucker-Smith:2001myb}. Searches designed for that possibility have followed (e.g.,~\cite{XENON:2017fdd} for one of the earliest examples).

The LUX-ZEPLIN (LZ) collaboration have extended its search window to approximately $270\keV$, motivated by these models~\cite{LZ:2026axp}. 
In $2.84\tyr$ of exposure they report a single event at $\ER = 248\pm23({\rm stat})\pm23({\rm sys})\keV$, in a region where the background expectation is $0.0106\pm0.0008$ counts. 
A profile likelihood ratio test disfavours the background-only hypothesis at a maximum local significance of $3.4\sigma$ across the models tested, 
falling to $2.6\sigma$ once the look-elsewhere effect over $293$ distinguishable signal spectra is included.

Interpretations have already begun to appear. LZ themselves work within an effective field theory framework~\cite{Fan:2010gt,Fitzpatrick:2012ix}, in which momentum-suppressed interactions push the signal to higher energy. Absorption models offer a different route~\cite{Lou:2026idn}, in which a sharply peaked signal is broadened by the energy resolution. Specific realisations proposed include the higgsino, one of the long-standing canonical dark matter candidates, whose two nearly degenerate neutral states supply the splitting~\cite{Fan:2026kxx,Freese:2026sga, Pospelov:2026ewn}.
 
In inelastic models the dark matter upscatters to a state heavier by $\delta$ during the collision, which raises the velocity required to produce a given recoil and removes the low-energy part of the spectrum. LZ scan $\delta$ from zero to $350\keV$ and find that the significance rises monotonically with the splitting, reaching its maximum at the largest value tested. Their tabulated significances are presented without comment, and we take up that trend here.

We ask what the LZ event would have to be if it is inelastic dark matter, and whether the interpretation can be tested. We first set out the rate calculation and the detector response we assume, then characterise the nuclear response in the extended window and identify the two diffraction minima from nuclear structure that bracket it. 
A likelihood fit in the mass--splitting plane shows that this structure forces the splitting upward, while the halo escape speed caps it from above.
We compute the annual modulation this implies, note a preliminary isotopic signature of the same splitting, and give the number of events a successor experiment needs to establish the modulation.

\begin{figure*}[t!]
\includegraphics[width=0.45 \textwidth]{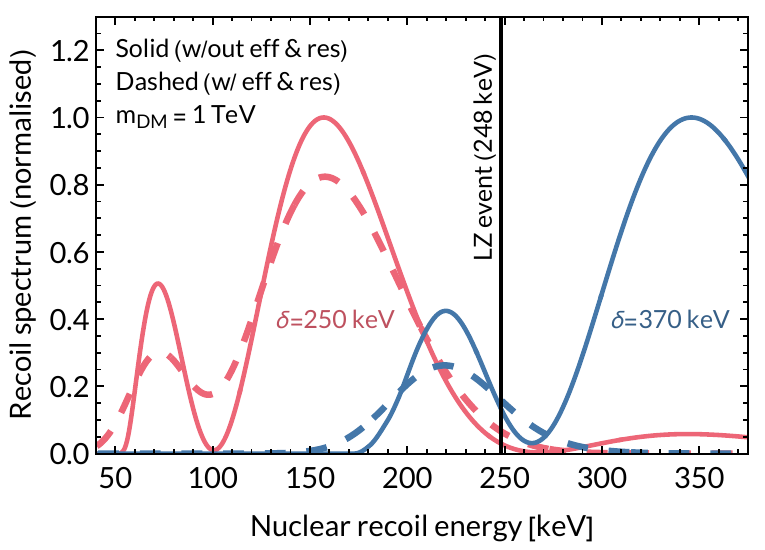}
\includegraphics[width=0.45 \textwidth]{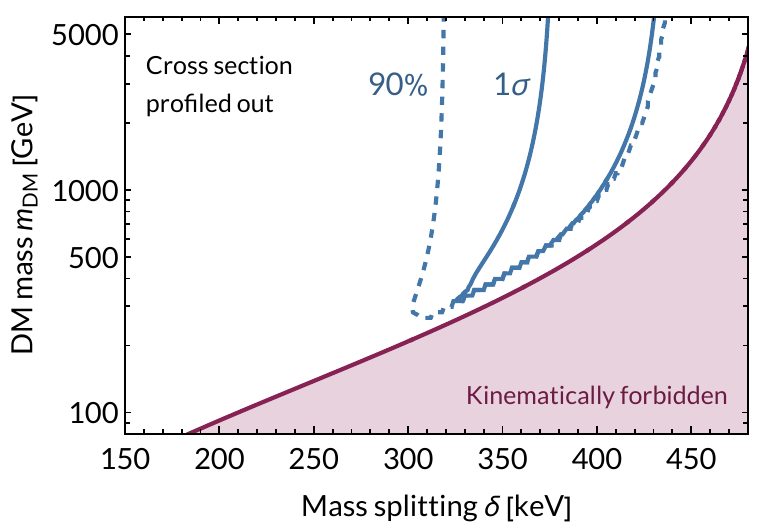}
\caption{\label{fig:spectra}
Left panel: spectra at $\mchi = 1\TeV$ for two mass splittings. 
Solid shows the spectrum before the detector response while dashed shows the result after applying the LZ efficiency in true recoil energy and smearing with eq.~\eqref{eq:smeared}. 
Each spectrum pair at fixed $\delta$ shares a common normalisation. The vertical line marks the observed event.
Right panel: confidence regions in the mass-splitting plane from eq.~\eqref{eq:loglike} with
the cross section profiled. The shaded region is kinematically inaccessible. 
The regions remain open towards large mass. 
For $\delta=250\keV$ the spectrum peaks near $150\keV$, below the observed event, which is why the fit disfavours small splittings.}
\end{figure*}

\shortsection{Scattering rate and parameter estimation} 
The differential event rate for both elastic and inelastic scattering of dark matter with a xenon nucleus of mass $m_A$ in the detector frame may be written as
\begin{equation}
\label{eq:dRdE}
\frac{d R}{d \ER}=\frac{1}{m_A} \frac{\rhoDM}{\mDM} \int_{\vmin} d^3v \, v f_{\rm{DM}}(\vec{v}+\vec{v}_{\rm{E}}) \frac{d \sigma}{d \ER} \;,
\end{equation}
where $\ER$ is the recoil energy of the xenon nucleus, $\mDM$ is the dark matter mass, $\rhoDM=0.3~\mathrm{GeV}/\mathrm{cm}^3$ is the canonical value of the local dark matter density~\cite{Read:2014qva}, $v$ and $\vec{v}$ are the dark matter speed and velocity, and~$f_{\rm{DM}}(\vec{v})$ is the dark matter velocity distribution in the galactic frame. 
We assume the isothermal Standard Halo Model, so that~$f_{\rm{DM}}(\vec{v})\propto\exp{\left(-v^2/v_0^2 \right)}$ is a Maxwell-Boltzmann distribution in the galactic frame with a hard cut-off at the galactic escape speed~$v_{\rm{esc}}$. 
We take~$v_{\rm{esc}}=544~\mathrm{km}/\mathrm{s}$ and a solar circular speed~$v_0=238~\mathrm{km}/\mathrm{s}$, following the recommendations of ref.~\cite{Baxter:2021pqo} (also adopted by LZ). 
We boost to the detector rest frame with $\vec{v}_{\rm{E}}=(0,v_0,0)+\vec{v}_{\rm{pec}}+\vec{v}_{\rm{e}}$, where $\vec{v}_{\rm{pec}}=(11.1,12.24,7.25)~\mathrm{km}/\mathrm{s}$~\cite{Schoenrich:2009bx} and we use the full expression for the Earth's velocity~$\vec{v}_{\rm{e}}$ from~\cite{McCabe:2013kea}, retaining the eccentricity of the orbit and the separate phases of the galactic radial, tangential and polar components.  As we will show, the signal we are concerned with sits close to the halo cut-off so for some parameters, the rate varies dramatically across the year. As a result, a careful characterisation of the Earth's velocity is needed. On the date of the LZ event the largest speed available in the laboratory frame is $v_{\rm max} = v_{\rm esc} + v_{\rm E} \approx 810\kms$, against a range of $781$ to $810\kms$ over the run.

The minimum speed to recoil with an energy $\ER$ additionally depends on the mass splitting~$\delta$ between two dark matter states:
\begin{equation}
\label{eq:vmin}
\vmin(\ER) = \frac{1}{\sqrt{2 m_A \ER}}\left(\frac{m_A \ER}{\mu_A} + \delta\right)\;,
\end{equation}
where~$\mu_{A}$ is the nucleus-dark matter reduced mass. The minimum speed is larger for bigger~$\delta$ for the simple reason that part of the kinetic energy of the incoming dark matter particle is required to excite the nucleus.

We work with the coherent spin-independent channel in what follows (as motivated by models such as the higgsino interpretation).
We therefore parameterise $d \sigma/d \ER$ in the canonical fashion (see e.g.~\cite{McCabe:2010zh}) in which the quantity is proportional to the square of the mass number ($A_N$) of the nucleus and a nuclear structure factor (or form factor) is needed. We sum eq.~\eqref{eq:dRdE} over the most dominant xenon isotopes for which shell-model structure factors are tabulated~\cite{Vietze:2014vsa}, weighting by natural abundance and carrying each isotope's own nuclear mass.

For a detector with energy resolution $\Delta E_\text{R} = 1.46\sqrt{E_\text{R}/\keV}\keV$, estimated from LZ's quoted statistical uncertainty of $23\keV$ at $248\keV$, the expected rate differential in reconstructed energy is
\begin{equation}
\mathcal{R}(E_\text{rec}) = \text{Exp} \int
\frac{e^{-(E_\text{rec}-E_\text{R})^2 / 2\Delta E_\text{R}^2}}{\sqrt{2\pi}\,\Delta E_\text{R}}
\, \epsilon(E_\text{R}) \, \frac{\text{d}R}{\text{d}E_\text{R}} \, \text{d}E_\text{R} \;,
\label{eq:smeared}
\end{equation}
where $\text{Exp}$ is the exposure and $\epsilon(E_\text{R})$ the published LZ efficiency curve, which falls to $50\%$ at $269.9\keV$. We apply the efficiency to the true recoil energy before smearing. 
 We evaluate $\mathcal{R}$ on a grid of $0.5\keV$ in true recoil energy.
 
 Every analysis in this work is in the window $125$ to $400\keV$ in reconstructed energy, chosen to be approximately free of background. 
 LZ report negligible electron-recoil leakage above $S1c = 250$~phd, which we convert to a recoil energy using their quoted $g_1 = 0.110$~phd and $g_2 = 34.5$~phd per electron. 
 Above this threshold the event at $248\keV$ is the only one in the data.
Nuclear-recoil backgrounds sum to at most $0.29$ events across LZ's full region of interest and are correspondingly negligible here, while accidental coincidences ($2.7\pm0.6$) are concentrated at low energies, leaving only a small fraction of one event above $125\keV$. 
Neither background can be represented in a signal-only likelihood below threshold, which is why we do not extend the window downward.

We now have all elements to calculate the recoil spectrum.  
Two such examples are shown in the left panel of fig.~\ref{fig:spectra} for two values of $\delta$ at $\mchi = 1\TeV$, normalised to a common peak height. 
The solid curves show the spectrum before the LZ efficiency and resolution are applied. 
In contrast, the dashed curves, kept at the same normalisation, show that their effect is to lower the peak and push it outward. 
The efficiency suppresses the spectrum at energies above $270$~keV. 
The diffraction minima of the xenon structure functions near $100$ and $250\keV$ are visible in both and further suppress the rate there.

The right panel shows the resulting confidence regions in the $(\mchi,\delta)$ plane, obtained from the extended profile likelihood of eq.~\eqref{eq:loglike}, with the cross section profiled in closed form since it enters the rate linearly. 
For $n$ observed events (here $n=1$) at energies $E_i$ and times $t_i$,
\begin{equation}
\ln \mathcal{L} = - n \ln \nu + \sum_i \ln \mathcal{R}(E_i, t_i) \;,
\label{eq:loglike}
\end{equation}
where $\mathcal{R}$ is the smeared rate of eq.~\eqref{eq:smeared} and $\nu = N(125\keV,
400\keV)$ is its integral over the analysis window and the exposure. 
As usual, terms that depend only on $n$ have been dropped from eq.~\eqref{eq:loglike}.
The exposure and the
reference cross section cancel identically between the two terms, so
eq.~\eqref{eq:loglike} is a pure shape fit: it asks where in energy and time the model places
the events, and carries no information about the absolute rate.
The arrival time is data, so the density is evaluated on the date of the event, while $\nu$ is integrated across the run,
which spans 27 March 2023 to 1 April 2024, with live time weighted uniformly.

At $\mchi = 1\TeV$ the splitting is confined to approximately $370$ to $400\keV$ at one standard deviation
and $350$ to $405\keV$ at $90\%$ confidence.
Small splittings are disfavoured because they place the spectrum at lower energy, while the observed event sits at $248\keV$. 
The regions are open towards large mass and cannot be closed by extending the scan. 
Fixing the cross section instead of profiling it can result in a closed region. 
With only one event, the asymptotic approximations underlying these intervals may not hold, and the preference for large $\delta$ should be read as indicative rather than fully robust statistical statements.

Also shown in the right panel of fig.~\ref{fig:spectra} is a kinematic constraint. 
The largest splitting the halo can supply is
\begin{equation}
\delta_{\rm max} = \tfrac{1}{2}\mu_A\, v_{\rm max}^2 ,
\qquad v_{\rm max} = \vesc + v_E ,
\label{eq:deltamax}
\end{equation}
with $v_E$ the Earth's speed through the halo. Read in the other direction,
eq.~\eqref{eq:deltamax} bounds the dark matter mass from below rather than selecting a
preferred value. Splittings of $250$, $300$ and $350\keV$ require masses above $149$, $234$
and $396\GeV$ respectively, rising to $817\GeV$ at $400\keV$.

\begin{figure}
\includegraphics[width=0.45 \textwidth]{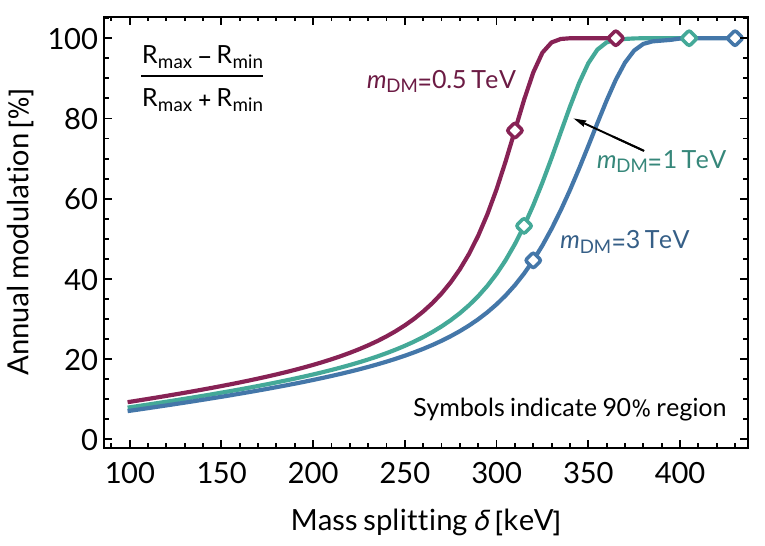}\\
\includegraphics[width=0.45 \textwidth]{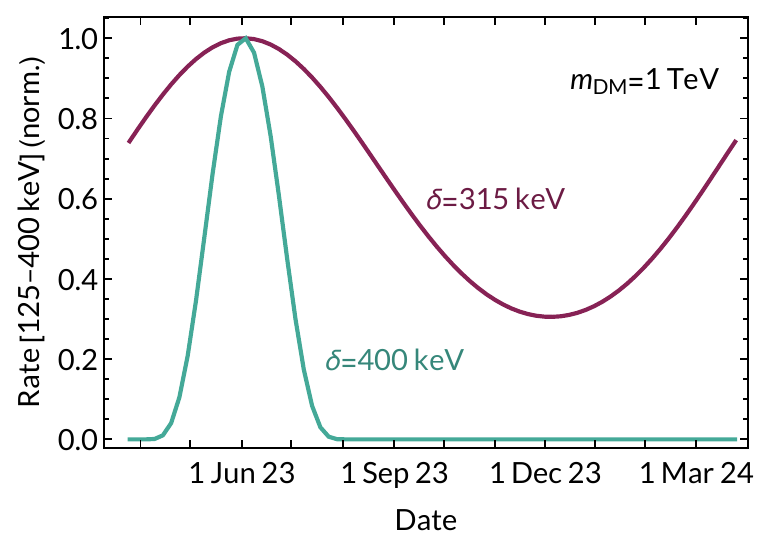}
\caption{\label{fig:mod}
Upper panel: the modulation amplitude $A = (R_{\rm max}-R_{\rm min})/(R_{\rm max}+R_{\rm min})$ of the rate in the $125$ to $400\keV$ window against mass splitting, for three dark matter
masses. 
Each curve stops where that mass can no longer produce a recoil at any time of year.
The region between the square symbols brackets the range identified by the fit in fig.~\ref{fig:spectra}. 
Lower panel: the rate through the year at $\mchi = 1\TeV$ for two splittings. 
At $315\keV$ the December rate is $0.31$ of the June peak, while at $400\keV$ the rate is zero for approximately two-thirds of the year, so the annual curve is not a small oscillation about a mean but peaks in June.
}
\end{figure}

\shortsection{Modulation}
As the fit in fig.~\ref{fig:spectra} shows, the surviving range of splittings lies close to the kinematic limit. We now explore what that implies. 
Taking $\delta = 350\keV$, $\mchi = 1\TeV$ and $\ER = 248\keV$ in eq.~\eqref{eq:vmin} we obtain $\vmin = 766\kms$, against $v_{\rm max} = 810\kms$ (in the detector frame) on the date of the event. 
The entire signal is supplied by dark matter lying within $44\kms$ of the escape speed, a sliver of the velocity distribution and the part of the local halo that is most uncertain (see e.g.,~\cite{McCabe:2010zh, OHare:2018trr, Besla:2019xbx, Folsom:2025lly}).

With such large values of $\delta$ favoured, other direct detection experiments will find it difficult to test the interpretation directly. 
The kinematic reach of eq.~\eqref{eq:deltamax} grows with the target's mass number $A_N$, since $\mu\to m_A$ for heavy dark matter, so no lighter target can supply a comparably large splitting. 
For example, argon is kinematically barred above $\delta = 131\keV$, only a target as heavy as xenon reaches sufficiently large $\delta_{\rm max}$.
While some candidates are available, notably tungsten in CRESST~\cite{CRESST:2019jnq}, they may find it challenging to scale to the exposure needed to observe the signal.

However, there is one potentially distinctive aspect of the signal that may aid confirmation of a dark matter origin in a xenon experiment. 
A signal supplied by a $44\kms$ sliver of the velocity distribution samples only its steep tail, so the Earth's orbital motion produces a large fractional change
in rate over the year. Such a possibility was also noted in ref.~\cite{Su:2026rwz, Nomura:2026qyq,DiMauro:2026ldr}. We define the modulation amplitude $A = (R_{\rm max}-R_{\rm min})/(R_{\rm max}+R_{\rm min})$, with the maximum and minimum taken over one year and $R$ the expected number of events in the analysis window. For a sinusoidal rate this is the familiar ratio of the modulated to the unmodulated component, but the rate here is not sinusoidal, and we return to that below. We obtain $A = 94\%$ for $\delta = 350\keV$ and $\mchi = 1\TeV$, against a few percent for elastic scattering in the same window. That inelastic scattering enhances modulation is a long-standing result. What is specific to this case is the magnitude, which follows from the high energy of the observed event pushing the splitting against the kinematic limit.

The upper panel of fig.~\ref{fig:mod} shows the modulation amplitude, expressed as a percentage, as a function of $\delta$ for three representative masses. 
While the recoil spectra are degenerate in mass around the TeV scale, the modulation is not. 
At $\delta = 350\keV$ it falls from $100\%$ at $500\GeV$ through $94\%$ at $1\TeV$ to $73\%$ at
$3\TeV$, so it is an observable that carries information about the mass in this regime. 
All three curves reach unity before the kinematic ceiling for that mass, beyond which the rate vanishes identically throughout the year.

The signal is not merely modulated but genuinely seasonal: present through the summer months and, for the largest splittings, absent altogether in winter.
An amplitude of unity means that the rate falls to exactly zero for part of the year. This is demonstrated in the lower panel of fig.~\ref{fig:mod}, which shows the integrated signal in
the range $125$ to $400\keV$ across one year at $\mchi = 1\TeV$ for two splittings. A sinusoidal description is a good approximation for $\delta = 315\keV$, but for
$\delta = 400\keV$ the rate is identically zero for approximately two-thirds of the year because $v_{\rm max}(t)$, driven by the Earth's orbital motion, falls below $\vmin$ altogether.

\shortsection{Reach with a larger dataset} 
To quantify what a measurement requires we use the Asimov formalism of ref.~\cite{Cowan:2010js}, applied to a nested test.
We write the rate as
\begin{equation}
\mathcal{R}(E_\text{rec}, t; \alpha) = \bar{\mathcal{R}}(E_\text{rec}) + \alpha\left[\mathcal{R}(E_\text{rec},t) - \bar{\mathcal{R}}(E_\text{rec})\right] ,
\label{eq:alpha}
\end{equation}
with $\bar{\mathcal{R}}$ the rate averaged over the year, so that $\alpha = 0$ is a signal with the same
recoil spectrum and no time dependence, and $\alpha = 1$ is the prediction with full time dependence. 
The null hypothesis is $\alpha = 0$, and the test statistic is the profile likelihood ratio
$q_0 = -2\ln\left[\mathcal{L}(\alpha=0)/\mathcal{L}(\hat\alpha)\right]$, which for one parameter of interest follows one half of a $\chi^2$ distribution with one degree of freedom,
so that the significance is $Z = \sqrt{q_0}$. Evaluating the likelihood on the Asimov dataset
built from $\alpha = 1$ gives 
\begin{equation}
q_{0,A} = 2\int \! \mathrm{d}E_\text{rec}\,\mathrm{d}t \left[ \mathcal{R}_1 \ln\frac{\mathcal{R}_1}{\mathcal{R}_0}
- \left(\mathcal{R}_1 - \mathcal{R}_0\right) \right] ,
\label{eq:asimov}
\end{equation}
where $\mathcal{R}_1$ is the modulated rate and $\mathcal{R}_0$ the best fit to it at $\alpha = 0$. 
Because the cross section is profiled in both hypotheses, the two predict the same total number of events, and the second term vanishes identically. 
What survives is proportional to that number, so the natural output is a count of events rather than an exposure.

\begin{figure}[t!]
\includegraphics[width=0.45 \textwidth]{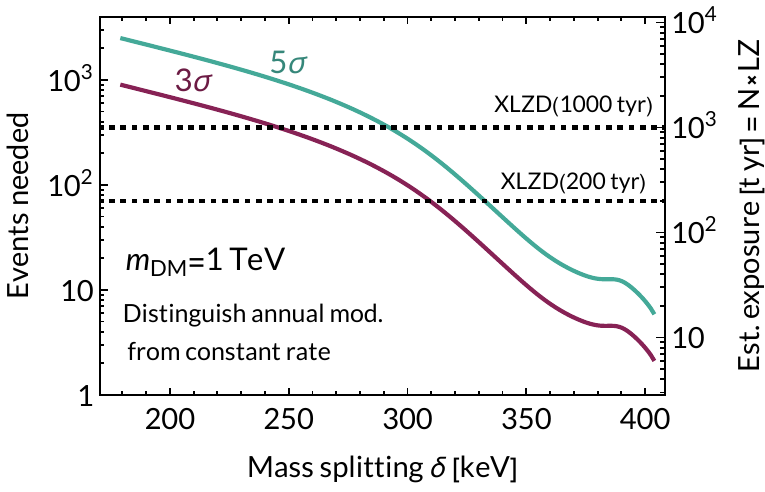}
\caption{\label{fig:distinguish}
Events required to distinguish an annually modulating signal from a constant rate, at $\mchi = 1\TeV$, for $3\sigma$ and $5\sigma$ median expected significance. 
We show the median expected significance obtained from the Asimov dataset of eq.~\eqref{eq:asimov}, with the cross section profiled. 
The statistic depends only on the distribution of events in reconstructed energy and arrival time. 
The right-hand axis converts the event count to an exposure at the rate LZ's single event implies, one event per $2.84\tyr$. 
This conversion carries a Poisson uncertainty of order its own size, so should be taken as a figure of merit only. Dotted lines mark two XLZD exposures.}
\end{figure}

Both hypotheses are signal: the test presumes the events are dark matter and asks only whether their arrival times carry the predicted structure, rather than testing for a signal
against background. 
A flat-in-time population would drive $\hat\alpha$ towards zero and falsify the interpretation rather than merely failing to confirm it. 
The test is also almost purely one of arrival times: repeating the calculation with the recoil energy of each event replaced by a
value drawn from the year-averaged spectrum leaves $q_{0,A}$ at $97.2\%$ of its full value at $\delta = 350\keV$, so only $2.8\%$ of the discriminating power comes from the correlation
between energy and arrival time. We also profile the mass and the splitting in the null, which is the conservative choice: 
it allows the unmodulated hypothesis to reshape its recoil spectrum to match the data
as closely as it can. This changes $q_{0,A}$ by less than one part in $10^6$, because no choice of those parameters can manufacture a time dependence, since the best an unmodulated hypothesis can do is average the true rate over the year, so the statistical power is irreducible.

Since $q_{0,A}$ scales linearly with the number of events $N$, we write $q_{0,A} = N\, q_1(\delta)$ with $q_1$ the value for a single event, so that the number of events needed for a target significance $Z$ is $N = Z^2/q_1(\delta)$. 
Scanning $\delta$ at $\mchi = 1\TeV$ and evaluating this at $Z=3$ and $Z=5$ gives the two curves of fig.~\ref{fig:distinguish}, with the requirement falling by two orders of magnitude across the
range of splittings shown. 
At the benchmark $\delta = 350\keV$, $3\sigma$ is reached at $11$ events and $5\sigma$ at $31$.

Normalising to the rate implied by LZ's single event, a figure of merit rather than a measured rate and one that carries a Poisson uncertainty of order its own size, an exposure of
$200\tyr$ delivers about $70$ events and a median significance of $7.5\sigma$ on the presence of modulation, while $1000\tyr$ delivers around $350$ events and $17\sigma$ on the same test.
These two exposures are marked in fig.~\ref{fig:distinguish} as benchmarks for XLZD~\cite{XLZD:2024nsu}, the
proposed successor to LZ. A next-generation detector~\cite{PANDA-X:2024dlo} is also planned within the PandaX programme and will achieve larger exposures than LZ.
LZ's own full exposure, of order $15\tyr$, yields around five events, already reaching $3\sigma$ for splittings above about $374\keV$, so within the range favoured by the fit.

\shortsection{Isotopic composition} 
 A second handle exists within xenon itself, and it is kinematic rather than nuclear. 
 Because $\delta_{\rm max}$ scales with the reduced mass, the lighter isotopes reach their ceilings first, so the isotopic composition of the signal is itself a function of the splitting.
 At $\mchi = 1\TeV$ we estimate that $^{136}$Xe supplies $9.7\%$ of the rate at $\delta = 250\keV$, close to its natural abundance of $8.9\%$, rising to $14.7\%$ at $350\keV$, $47\%$ at $380\keV$ and all of it at $400\keV$.

We note this effect rather than propose it as a practical measurement, since enrichment to the levels that would make it a useful probe of $\delta$ is not realistic at the scale a dark
matter search requires. 
Even so, the direction is informative: a target enriched in $^{136}$Xe would gain rate relative to a natural target as $\delta$ rises, while one enriched in $^{129}$Xe would lose it, so the sign of any such difference indicates where $\delta$ sits relative to $\delta_{\rm max}$. 

This probes $\delta/\delta_{\rm max}$ rather than $\delta$ itself, as the modulation amplitude does, so the two are not independent measurements of different quantities. 
Their systematics are independent, however, the modulation working through the time dependence of $v_{\rm max}$ and the isotopic composition through its dependence on the target mass, so they
constitute a cross-check rather than a coincidence.

\shortsection{Summary and outlook}
A single event at $248\keV$, if interpreted as inelastic dark matter, favours a splitting close to the kinematic limit allowed by the speed of dark matter in the halo. 
A small splitting would place most of the signal below the observed energy, where an excess of events is not observed, 
so the fit is pushed towards large $\delta$ until the halo escape speed caps it from above.

A splitting this close to the kinematic limit makes the signal genuinely seasonal:
the annual modulation rises above $50\%$ as $\delta$ moves into the preferred parts of parameter space identified in this work, even reaching $100\%$ for
some parameters, meaning that for part of the year no dark matter in the halo has enough
speed to scatter at all. 
We have characterised the number of events needed to establish this modulation, under the assumptions of the Standard Halo Model. 
XLZD and PandaX, with target masses larger than LZ's, could exploit this directly. 
Ensuring that experiments record data in June, when the dark matter rate peaks, would also help avoid missing this or a similar signal that is more prominent in summer.

Before ending, we comment that all of our results rest on the Standard Halo Model. 
The signal is supplied by the tail of the velocity distribution nearest the escape speed, the part about which the model is most uncertain, so how deviations from it would change these results is a
question we leave for future work.

\shortsection{Acknowledgments}
I acknowledge funding support from the STFC. 
I am grateful to CERN for hospitality while this work was carried out, and I am particularly happy to acknowledge fruitful discussions with Miguel Escudero, David Cerdeno and Matthew McCullough. 
Claude Opus 5.0 was used to validate some of the calculations in this work. I take responsibility for any mistakes that remain.

\bibliography{refs}

@article{Su:2026rwz,
    author = "Su, Liangliang and Yang, Jin Min and Yang, Wen-Na",
    title = "{Inelastic Dark Matter Signature at High Recoil Energy in LUX-ZEPLIN and CRESST}",
    eprint = "2609.01475",
    archivePrefix = "arXiv",
    primaryClass = "hep-ph",
    month = "9",
    year = "2026"
}

@article{Nomura:2026qyq,
    author = "Nomura, Yasunori",
    title = "{Dark Matter as the Z{\_}2 Partner of the Standard Model Higgs Boson}",
    eprint = "2609.02505",
    archivePrefix = "arXiv",
    primaryClass = "hep-ph",
    reportNumber = "RIKEN-iTHEMS-Report-26",
    month = "9",
    year = "2026"
}

@article{OHare:2018trr,
    author = "O'Hare, Ciaran A. J. and McCabe, Christopher and Evans, N. Wyn and Myeong, GyuChul and Belokurov, Vasily",
    title = "{Dark matter hurricane: Measuring the S1 stream with dark matter detectors}",
    eprint = "1807.09004",
    archivePrefix = "arXiv",
    primaryClass = "astro-ph.CO",
    doi = "10.1103/PhysRevD.98.103006",
    journal = "Phys. Rev. D",
    volume = "98",
    number = "10",
    pages = "103006",
    year = "2018"
}

@article{Besla:2019xbx,
    author = "Besla, Gurtina and Peter, Annika and Garavito-Camargo, Nicolas",
    title = "{The highest-speed local dark matter particles come from the Large Magellanic Cloud}",
    eprint = "1909.04140",
    archivePrefix = "arXiv",
    primaryClass = "astro-ph.GA",
    doi = "10.1088/1475-7516/2019/11/013",
    journal = "JCAP",
    volume = "11",
    pages = "013",
    year = "2019"
}

@article{Folsom:2025lly,
    author = "Folsom, Dylan and Blanco, Carlos and Lisanti, Mariangela and Necib, Lina and Vogelsberger, Mark and Hernquist, Lars",
    title = "{Dark Matter Velocity Distributions for Direct Detection: Astrophysical Uncertainties Are Smaller Than They Appear}",
    eprint = "2505.07924",
    archivePrefix = "arXiv",
    primaryClass = "hep-ph",
    doi = "10.1103/wmpq-mw4h",
    journal = "Phys. Rev. Lett.",
    volume = "135",
    number = "21",
    pages = "211004",
    year = "2025"
}

@article{PANDA-X:2024dlo,
    author = "Abdukerim, Abdusalam and others",
    collaboration = "PANDA-X, PandaX",
    title = "{PandaX-xT{\textemdash}A deep underground multi-ten-tonne liquid xenon observatory}",
    eprint = "2402.03596",
    archivePrefix = "arXiv",
    primaryClass = "hep-ex",
    doi = "10.1007/s11433-024-2539-y",
    journal = "Sci. China Phys. Mech. Astron.",
    volume = "68",
    number = "2",
    pages = "221011",
    year = "2025"
}

@article{DiMauro:2026ldr,
    author = "Di Mauro, Mattia",
    title = "{Dark Matter at the Kinematic Edge: Interpreting the 248 keV LZ Nuclear-Recoil Candidate}",
    eprint = "2609.02608",
    archivePrefix = "arXiv",
    primaryClass = "hep-ph",
    month = "9",
    year = "2026"
}

@article{XLZD:2024nsu,
    author = "Aalbers, J. and others",
    collaboration = "XLZD",
    title = "{The XLZD Design Book: towards the next-generation liquid xenon observatory for dark matter and neutrino physics}",
    eprint = "2410.17137",
    archivePrefix = "arXiv",
    primaryClass = "hep-ex",
    doi = "10.1140/epjc/s10052-025-14810-w",
    journal = "Eur. Phys. J. C",
    volume = "85",
    number = "10",
    pages = "1192",
    year = "2025"
}

@article{Cowan:2010js,
    author = "Cowan, Glen and Cranmer, Kyle and Gross, Eilam and Vitells, Ofer",
    title = "{Asymptotic formulae for likelihood-based tests of new physics}",
    eprint = "1007.1727",
    archivePrefix = "arXiv",
    primaryClass = "physics.data-an",
    doi = "10.1140/epjc/s10052-011-1554-0",
    journal = "Eur. Phys. J. C",
    volume = "71",
    pages = "1554",
    year = "2011",
    note = "[Erratum: Eur.Phys.J.C 73, 2501 (2013)]"
}

@article{CRESST:2019jnq,
    author = "Abdelhameed, A. H. and others",
    collaboration = "CRESST",
    title = "{First results from the CRESST-III low-mass dark matter program}",
    eprint = "1904.00498",
    archivePrefix = "arXiv",
    primaryClass = "astro-ph.CO",
    doi = "10.1103/PhysRevD.100.102002",
    journal = "Phys. Rev. D",
    volume = "100",
    number = "10",
    pages = "102002",
    year = "2019"
}

@article{McCabe:2010zh,
    author = "McCabe, Christopher",
    title = "{The Astrophysical Uncertainties Of Dark Matter Direct Detection Experiments}",
    eprint = "1005.0579",
    archivePrefix = "arXiv",
    primaryClass = "hep-ph",
    doi = "10.1103/PhysRevD.82.023530",
    journal = "Phys. Rev. D",
    volume = "82",
    pages = "023530",
    year = "2010"
}

@article{Vietze:2014vsa,
    author = "Vietze, L. and Klos, P. and Men{\'e}ndez, J. and Haxton, W. C. and Schwenk, A.",
    title = "{Nuclear structure aspects of spin-independent WIMP scattering off xenon}",
    eprint = "1412.6091",
    archivePrefix = "arXiv",
    primaryClass = "nucl-th",
    doi = "10.1103/PhysRevD.91.043520",
    journal = "Phys. Rev. D",
    volume = "91",
    number = "4",
    pages = "043520",
    year = "2015"
}

@article{McCabe:2013kea,
    author = "McCabe, Christopher",
    title = "{The Earth's velocity for direct detection experiments}",
    eprint = "1312.1355",
    archivePrefix = "arXiv",
    primaryClass = "astro-ph.CO",
    reportNumber = "IPPP-13-95, DCPT-13-190",
    doi = "10.1088/1475-7516/2014/02/027",
    journal = "JCAP",
    volume = "02",
    pages = "027",
    year = "2014"
}

@article{Schoenrich:2009bx,
    author = "Schoenrich, R. and Binney, J. and Dehnen, W.",
    title = "{Local Kinematics and the Local Standard of Rest}",
    eprint = "0912.3693",
    archivePrefix = "arXiv",
    primaryClass = "astro-ph.GA",
    doi = "10.1111/j.1365-2966.2010.16253.x",
    journal = "Mon. Not. Roy. Astron. Soc.",
    volume = "403",
    pages = "1829",
    year = "2010"
}

@article{Baxter:2021pqo,
    author = "Baxter, D. and others",
    title = "{Recommended conventions for reporting results from direct dark matter searches}",
    eprint = "2105.00599",
    archivePrefix = "arXiv",
    primaryClass = "hep-ex",
    doi = "10.1140/epjc/s10052-021-09655-y",
    journal = "Eur. Phys. J. C",
    volume = "81",
    number = "10",
    pages = "907",
    year = "2021"
}

@article{Read:2014qva,
    author = "Read, J. I.",
    title = "{The Local Dark Matter Density}",
    eprint = "1404.1938",
    archivePrefix = "arXiv",
    primaryClass = "astro-ph.GA",
    reportNumber = "JPHYSG-100038.R1",
    doi = "10.1088/0954-3899/41/6/063101",
    journal = "J. Phys. G",
    volume = "41",
    pages = "063101",
    year = "2014"
}

@article{Lou:2026idn,
    author = "Lou, Yuanchao and Lu, Chih-Ting",
    title = "{Fermionic Dark Matter Absorption and the High-Energy Event in LUX-ZEPLIN}",
    eprint = "2609.01592",
    archivePrefix = "arXiv",
    primaryClass = "hep-ph",
    month = "9",
    year = "2026"
}

@article{Pospelov:2026ewn,
    author = "Pospelov, Maxim and Ramani, Harikrishnan",
    title = "{Strong Constraints on Higgsino Dark Matter from Solar Capture}",
    eprint = "2609.02775",
    archivePrefix = "arXiv",
    primaryClass = "hep-ph",
    month = "9",
    year = "2026"
}

@article{Freese:2026sga,
    author = "Freese, Katherine and Theodosopoulos, Dionysios P.",
    title = "{Higgsino Dark Matter Interpretation of the LUX-ZEPLIN 248 keV Nuclear-Recoil Event}",
    eprint = "2609.01583",
    archivePrefix = "arXiv",
    primaryClass = "hep-ph",
    month = "9",
    year = "2026"
}

@article{Fan:2026kxx,
    author = "Fan, JiJi and Reece, Matthew",
    title = "{Higgsino Above the Sea of Fog}",
    eprint = "2609.01504",
    archivePrefix = "arXiv",
    primaryClass = "hep-ph",
    month = "9",
    year = "2026"
}

@article{Fitzpatrick:2012ix,
    author = "Fitzpatrick, A. Liam and Haxton, Wick and Katz, Emanuel and Lubbers, Nicholas and Xu, Yiming",
    title = "{The Effective Field Theory of Dark Matter Direct Detection}",
    eprint = "1203.3542",
    archivePrefix = "arXiv",
    primaryClass = "hep-ph",
    doi = "10.1088/1475-7516/2013/02/004",
    journal = "JCAP",
    volume = "02",
    pages = "004",
    year = "2013"
}

@article{Fan:2010gt,
    author = "Fan, JiJi and Reece, Matthew and Wang, Lian-Tao",
    title = "{Non-relativistic effective theory of dark matter direct detection}",
    eprint = "1008.1591",
    archivePrefix = "arXiv",
    primaryClass = "hep-ph",
    doi = "10.1088/1475-7516/2010/11/042",
    journal = "JCAP",
    volume = "11",
    pages = "042",
    year = "2010"
}

@article{LZ:2024zvo,
    author = "Aalbers, J. and others",
    collaboration = "LZ",
    title = "{Dark Matter Search Results from 4.2{\,}{\,}Tonne-Years of Exposure of the LUX-ZEPLIN (LZ) Experiment}",
    eprint = "2410.17036",
    archivePrefix = "arXiv",
    primaryClass = "hep-ex",
    reportNumber = "FERMILAB-PUB-24-0796-V",
    doi = "10.1103/4dyc-z8zf",
    journal = "Phys. Rev. Lett.",
    volume = "135",
    number = "1",
    pages = "011802",
    year = "2025"
}

@article{MarrodanUndagoitia:2015veg,
    author = "Marrod{\'a}n Undagoitia, Teresa and Rauch, Ludwig",
    title = "{Dark matter direct-detection experiments}",
    eprint = "1509.08767",
    archivePrefix = "arXiv",
    primaryClass = "physics.ins-det",
    doi = "10.1088/0954-3899/43/1/013001",
    journal = "J. Phys. G",
    volume = "43",
    number = "1",
    pages = "013001",
    year = "2016"
}

@article{Chang:2009yt,
    author = "Chang, Spencer and Pierce, Aaron and Weiner, Neal",
    title = "{Momentum Dependent Dark Matter Scattering}",
    eprint = "0908.3192",
    archivePrefix = "arXiv",
    primaryClass = "hep-ph",
    reportNumber = "MCTP-09-42",
    doi = "10.1088/1475-7516/2010/01/006",
    journal = "JCAP",
    volume = "01",
    pages = "006",
    year = "2010"
}

@article{Tucker-Smith:2001myb,
    author = "Tucker-Smith, David and Weiner, Neal",
    title = "{Inelastic dark matter}",
    eprint = "hep-ph/0101138",
    archivePrefix = "arXiv",
    reportNumber = "UCB-PTH-00-43, LBNL-47234, UW-PT-00-17",
    doi = "10.1103/PhysRevD.64.043502",
    journal = "Phys. Rev. D",
    volume = "64",
    pages = "043502",
    year = "2001"
}

@article{XENON:2017fdd,
    author = "Aprile, E. and others",
    collaboration = "XENON",
    title = "{Effective field theory search for high-energy nuclear recoils using the XENON100 dark matter detector}",
    eprint = "1705.02614",
    archivePrefix = "arXiv",
    primaryClass = "astro-ph.CO",
    doi = "10.1103/PhysRevD.96.042004",
    journal = "Phys. Rev. D",
    volume = "96",
    number = "4",
    pages = "042004",
    year = "2017"
}

@article{LZ:2026axp,
    author = "Akerib, D. S. and others",
    collaboration = "LZ",
    title = "{Search for dark matter particle interactions in an extended nuclear recoil energy window with the LUX-ZEPLIN (LZ) experiment}",
    eprint = "2609.02823",
    archivePrefix = "arXiv",
    primaryClass = "hep-ex",
    month = "9",
    year = "2026"
}
\bibliographystyle{apsrev4-1}

\end{document}